\documentclass[twocolumn,twoside,slac_two]{revtex4}
\usepackage{graphicx}
\usepackage{fancyhdr}
\renewcommand{\headrulewidth}{0pt}
\renewcommand{\footrulewidth}{0pt}

\newcommand{\be}[0]{\begin{equation}}
\newcommand{\ee}[0]{\end{equation}}
\newcommand{\ba}[0]{\begin{eqnarray}}
\newcommand{\ea}[0]{\end{eqnarray}}

\begin{document}

\title{Possible considerations to teleport fermionic particles via studying on teleportation of two-particle state with a four fermionic-particle pure entangled state

}


\author{Shilan. Savan}
\affiliation{Physics Department, Semnan University, Semnan, Iran}

\author{Mehrdad. Ghominejad}
\affiliation{Physics Department, Semnan University, Semnan, Iran }

\begin{abstract}
In this paper we have firstly recapped some evolutionary debates
on conceptual quantum information matters,followed by an
experiment done by Lamei-Rashti and his collaborator,by which the
bell inequality on p-p scattering is violated.We then, by using
the goal of his experiment, thought to arrange POVM formalism for
a possible teleportation of two particle states, via nuclear
magnetic spin of four entangled hydrogen like atoms.
\end{abstract}

\maketitle

\thispagestyle{fancy}


\section{Introduction}\label{section1}
The paradox of Einstein, Podolsky and Rosen \cite{A.
Einstein:1935} was advanced as an argument that quantum mechanics
could not be a complete theory but should be supplemented by
additional variables, by which the causality and locality theories
are restored. According to the local realistic theories, objects
should have definite properties whether they are measured or not
(reality), and there is no action-at a-distance in nature
(locality). Some attempts were made to explain quantum mechanical
phenomena from a view of the local realistic theories. Bell,
however, showed quantum mechanical correlations between entangled
systems, can be stronger than those by the local realistic
theories \cite{J. S. Bell:1964}. Since Bell's proof was given by
an inequality which could be tested experimentally[3] and [4]. The
mentioned movement to pursue and check experimentally the bell
inequality in QCD interactions started with the key paper of
Clauser, Horne, Shimony and Holt (CHSH) \cite{John F.
Clauser:1969} who generalized Bell's theorem such that it applies
to realizable experimental tests with pair detection of all local
hidden variable theories.Quantum mechanically thinking, the
mentioned inequality can not be totally true though. To check this
non reliability for photons and most importantly for QCD like
scatterings such as Proton-Proton (called P-P scattering) in low
and rarely for high energies to actually observe such a violation
of Bell's inequality in a laboratory some experiments was done.
most of the experiments to test Bell's inequality performed so far
have used spin correlations of a two photon system. The only one
exception is the experiment by Lamehi- Rachti and Mittig
(LRM)\cite{M. Lamehi-Rachti:1976} with protons in entangled
systems. They used strong interaction to test the Bell's
inequality. Since the strong interaction is a short range
interaction, entangled particles are produced with extremely short
coherence length. It is of considerable interest to investigate
whether an entanglement between two particles is robustly
maintained even if the two particles are spatially separated from
each other by a distance extremely beyond their coherence length.
Measured spin-correlations between two protons in the spin-singlet
 state which was produced by the proton-proton S-wave elastic
scattering. This is because proton-proton scattering at large
angle and low energy, say a few MeV, goes mainly in S wave But the
antisymmetry of the final wave function then requires the anti
symmetries singlet spin state. In this state, when one spin is
found 'up' the other is found 'down'. This follows formally from
the quantum expectation value,$<sin
glet|\sigma_{z}(1)\sigma_{z}(2)|singlet>=-1$ In these experiments,
protons of 14 MeV lab energy are scattered at a lab angle of 45°,
and spin correlation of scattered and recoil protons are measured.
This experiment anyhow, shows agree with quantum mechanics and
disagree with the locality inequality, and are the first serious
test of Bell inequality by spin=1/2 fermion system with mass.
Violating the Bell inequality for the sake of EPR states, opens up
the world of teleportation for us, that has recently been the
source of many researches in this domain.

\section{Quantum Teleportation}
Quantum teleportation is a technique for moving quantum states
around, even in the absence of a quantum communications channel
linking the sender of quantum state to the recipient. Now it seems
necessary to explain some crucial concepts, namely, entanglement,
qubit and Bell state, which have key roles to perceive
teleportation.
\subsection{Entanglement}
In order to understand the concept of quantum communication the
phenomena called entanglement, described for the first time at
1935 by Einstein, Podolsky and Rosen.the EPR interactions have
three main characteristics which make them unique and extremely
interesting:
\begin{itemize}
    \item Any modification applied to one of the particles, from now on
    particle 1, would imply a variation on its couple (particle 2).The
    alteration on particle 2 would depend on the modification done on
    particle1. Different modifications would imply different
    variations.
    \item The alteration of particle 2, due to the
    modification of particle 1 does not depend on the distance between
    Them.
    \item Particle 2 is instantly altered once particle 1 is
    modified
\end{itemize}
\subsection{Measurment-POVM}
Quantum measurments are described by a collection $\{M_{m}\}$of
measurement operators. These operators acting on the state space
of   the system being measured. The index m refers to the
measurement outcomes that may occur in the
experiment.Distinguishing quantum states is the important problem
in Quantum information and Quantum computation. In such instances
there is a mathematical tool known as the POVM \cite{Asher
Peres}(Positive Operator-Value Measure) formalism which is
especially well adapted to the analysis of the measurements. the
elements of a POVM are not necessarily orthogonal, with the
consequence that the number of elements in the POVM, n, can be
larger than the dimension, N, of the Hilbert space they act in.
Any measurement process can be described in terms of a quantum
operation in the following way:
\begin{enumerate}
    \item  To the set of possible outcomes $\{m\}$ from a measurement a set of Quantum
    operations $\{E_{m}\}$ is associated.
    \item  Each $E_{m}$ describes the dynamics of system when outcome m is found.
    \item  The probability $P_{m}$ of the outcome m is
    $Tr[E_{m}(\rho)]$ and the post-measurement state is given by
    $\frac{E_{m}(\rho)}{Tr[E_{m}(\rho)}$
    \item  The total Quantum operation $ E=\sum_{m}E_{m} $ is trace Preserving, because the probabilities
    $p_{m}$the distinct outcomes Sum to One.
\end{enumerate}
\textbf{POVM measurement} A set of Operators $\{E_{m}\}$satisfying
\begin{itemize}
    \item $ E_{m}$\hspace{5cm}          Positivity
    \item $\sum_{m}E_{m}=1 $\hspace{3.2cm}            Completness
    \item $ p_{m}=Tr[E_{m}\rho]=\langle\Psi|E_{m}|\Psi\rangle$\hspace{1cm}     Probability rule
\end{itemize}
Defining $M_{m}\equiv\sqrt{E_{m}}$ we see that
$\sum_{m}{M_{m}^{\dag}M_{m}}=\sum_{m}{E_{m}}=I$, and therefore the
set $\{M_{m\}}$ describes a measurement with
\textbf{POVM}$\{E_{m}\}$.

\subsection{Qubit}
just as a classical bit has a state,either 0 or 1 a qubit also has
a state, witch can be thought of as a vector in a two-dimensional
\emph{Hilbert}-space and will be denoted as$ |0> and |1>$ from now
on. the main and important difference between bits and qubits is
that the latter can also be in a linear combination of
states,i.e.a coherent superposion:$|\Psi>=\alpha|0>+\beta|1>$
,where $\alpha$ and $\beta$ are complex numbers.
 Qubit is a fundamental element for quantum
computation and quantum information. In the early days of quantum
mechanics the qubit structure was not at all obvious, and people
struggle with phenomena that we may now understand in terms of
qubits, that is, in the terms of two level quantum systems.One
very useful picture when thinking about qubits is the geometrical
representation of polarization states on the so-called
Blotch-sphere and often serves as an excellent testbed for ideas
about quantum computation and quantum information.one can rewrite
the state as
\begin{equation}
    |\Psi>=\cos(\Theta/2)|0>+e^{i\Phi}\sin(\Theta/2)|1>
\end{equation}. where the angle $\Theta$ and $\Phi$ define a point on
the three-dimensional unit sphere shown in
\subsection{Bell state}
For two classical bits there are four possible states, 00, 01, 10
and 11, but a pair of qubits can also exist in a superposition of
this states, therefore spanning a 4-dimensional Hilbert space. One
remarkable feature of such states is that they cannot be built as
single and separable qubit states$|a>$ and $|b>$ such that
$|\Phi>=|a>|b>$.Thus, they cannot be written as a product of
states of their component systems, which is a very crucial
property of entangled states.Einstein, Podolsky and Rosen pointed
out these strange properties of such states and they have been
named Bell-States in honour of John Bell, who showed that
correlations in such entangled states are stronger than could
possibly exist between classical systems. for a two-qubit system
there are four distinct entangled states, the Bell-States,
        $$|\Phi^{\pm}>=(1/\sqrt{2})(|00>\pm|11>)$$        $$|\Psi^{\pm}>=(1/\sqrt{2})(|01>\pm|10>)$$
which form an orthonormal basis for the two-qubit state space, and
can therefore be distinguished by appropriate quantum
measurements.
  The basic idea in quantum
teleportation is the following: Suppose we have two parties, Alice
and Bob. Say Alice wishes to transfer a certain quantum particle
to Bob, but cannot do so directly. According to the rules of
quantum mechanics if she measured the qubit this action would
destroy the quantum state of the particle without revealing her
all the necessary information which she could then send to Bob to
reconstruct the qubit.how Alice can provide Bob with her quantum
particle. The solution is once again entanglement. In outline, the
steps of the solution are as follows: Alice interacts the qubit
$|\Psi>$ with half of EPR pair (singlet state), and then measures
the two qubits in her possession, obtaining one of four possible
classical results, 00, 01, 10 and 11. She sends this information
to Bob. Depending on Alice's classical message, Bob performs one
of four operations on his half of the EPR pair. Amazingly, by
doing this he recovers the original state $|\Psi>!$
 The state to be eleported is $|\Psi>=\alpha|0>+\beta|1>$, where $\alpha$ and $\beta$ are
unknown amplitudes. the state input into the quantum circuit
$|\Psi_{0}>$ is:\begin{eqnarray}
 |\Psi_{0}>=(1/\sqrt{2})[\alpha|0>(|00>+|11>)\nonumber \\
 +\beta|1>(|00>+|11>)]
\end{eqnarray}

The first two qubits (on the left) belong to Alice, and the third
qubit to Bob. Alice sends her qubits through a CNOT gate, and then
sends the first qubit through a Hadamard gate,
obtaining,
\begin{eqnarray}
 |\Psi_{1}>=(1/(\sqrt{2})[\alpha|0>(|00>+|11>)\nonumber\\+\beta|1>(|10>+|01>)]
 \end{eqnarray}
\begin{eqnarray}
 |\Psi_{2}>=(1/\sqrt{2})[\alpha(|0>+|1>)(|00>+|11>)\nonumber\\
 +\beta(|0>-|1>)(|10>+|01>)]
\end{eqnarray}
The latter state may be re-written in the following way, simply by
regrouping terms:
$$|\Psi_{2}>=(1/\sqrt{2})[|00>(\alpha|0>+\beta|1>)+$$$$|01>(\alpha|1>+\beta|0>)+|10>(\alpha|0>-\beta|1>)$$$$+|11>(\alpha|1>-\beta|0>)]$$

Obviously Alice obtains one of the four possible two-bit results
among 00, 01, 10 or 11 as measurement outcome. Each of them is in
close connection with the state of Bob's qubit hence Alice sends
these two classical bits to Bob. After a short hesitation Bob
compares$|\Psi>$to the potential states of his half Bell pair. It
is easy to realize the
$00\rightarrow\frac{\alpha|0\rangle+\beta|1\rangle}{2}=I|\Psi\rangle$
$01\rightarrow\frac{\alpha|1\rangle+\beta|0\rangle}{2}=X|\Psi\rangle$
$10\rightarrow\frac{\alpha|0\rangle-\beta|1\rangle}{2}=Z|\Psi\rangle$
$11\rightarrow\frac{\alpha|1\rangle-\beta|0\rangle}{2}=ZX|\Psi\rangle$
Therefore Bob has only to apply the inverse of the appropriate
transform(s) in compliance with the received classical bits. \\ So
as it can be seen, quantum teleportation is a process of
transmission of an unknown quantum state via a previously shared
EPR pair with the help of only two classical bits transmitted
through a classical channel \cite{Bennett93}. It was regarded as
one of the most striking progress of quantum information theory
\cite{NielsenChuang}. Suppose that the sender Alice has two
particles 1,2 in an unknown state $|00\rangle$
\begin{equation}
|\Phi\rangle_{12}=(a|00\rangle+b|01\rangle+c|10\rangle+d|11\rangle)_{12},
\end{equation}
where $a, b, c, d$ are arbitrary complex numbers, and satisfy $|a|^2+|b|^2+|c|^2+|d|^2=1$. We also suppose that
Alice and Bob share  quantum entanglement in the form of
following  partly pure entangled four-particle state, which will
be used as the quantum channel,
\begin{equation}
|\Phi\rangle_{3456}=(\alpha|0000\rangle+\beta|1001\rangle+\gamma|0110\rangle+\delta|1111\rangle)_{3456},
\end{equation}
The particles 3 and 4, and particle pair (1, 2) are in Alice's
possession, and particles 5 and 6 are in Bob's possession. The
overall state of six particles is
\begin{equation}
|\Phi\rangle_w=|\Phi\rangle_{12}\otimes|\Phi\rangle_{3456}.
\end{equation}
In order to realize the teleportation, firstly Alice performs  two
Bell state measurements on particles 2,3 and 1,4, If the outcomes
of the Alice's two Bell state measurements
are$|\Phi^+\rangle_{23}$and$|\Phi^+\rangle_{14}$then the particle
5 and 6 are in the state,
\begin{eqnarray}
|\Psi_0\rangle_{56}=\frac{_{14}\langle\Phi^+|_{23}\langle\Phi^+|\Phi\rangle_w}{|_{14}\langle\Phi^+|_{23}\langle\Phi^+|\Phi\rangle_w|}
\nonumber \\
=\frac{1}{\sqrt{|a\alpha|^2+|b\beta|^2+|c\gamma|^2+|d\delta|^2}} \nonumber \\
(a\alpha|00\rangle+b\beta|01\rangle+c\gamma|10\rangle+d\delta|11\rangle)_{56},
\end{eqnarray}

Now, we will not write out the states of the particles 5 and 6
corresponding to the other outcomes of Alice's two Bell state
measurements\cite{YanTanYang}. Then Alice informs Bob her two Bell
state measurements on particles 2, 3 and 1, 4  Without loss of
generality,  we give the case for$|\Psi_0\rangle_{56}$all other
cases can be deduced similarly. In order to realize the
teleportation, Bob introduces two auxiliary qubits $a and b$ in
the state $|00\rangle_{ab}$. Thus the state of particles $5, 6, a,
and b$ becomes, $|\Psi_0\rangle_{56}|00\rangle_{ab}$ Then Bob
performs two controlled-not operations (CNOT gate) with particles
5 and 6 as the control qubits and the auxiliary particles $a and
b$ as the target qubits respectively. After completing this
operation the particles 5, 6, a, and b are in the following state,
\begin{eqnarray}
|\Psi'_0\rangle_{56ab}=\frac {1}{\sqrt
{|a\alpha|^2+|b\beta|^2+|c\gamma|^2+|d\delta|^2}}\nonumber
\\ (a\alpha|0000\rangle+b\beta|0101\rangle+
c\gamma|1010\rangle+d\delta|1111\rangle)_{56ab}.
\end{eqnarray}
After some rearrangement one obtains,

$$|\Psi'_0\rangle_{56ab}=\frac {1}{4\sqrt
{|a\alpha|^2+|b\beta|^2+|c\gamma|^2+|d\delta|^2}}$$
$$
(a|01\rangle+b|01\rangle+c|10\rangle+d|11\rangle)_{56}\otimes $$
$$(\alpha|00\rangle+\beta|01\rangle
+\gamma|10\rangle+\delta|11\rangle)_{ab}$$$$
+(a|00\rangle+b|01\rangle-c|10\rangle-d|11\rangle)_{56}\otimes$$
$$ (\alpha|00\rangle+\beta|01\rangle
-\gamma|10\rangle-\delta|11\rangle)_{ab}$$
$$+(a|00\rangle-b|01\rangle+c|10\rangle-d|11\rangle)_{56}\otimes$$
$$ (\alpha|00\rangle-\beta|01\rangle
+\gamma|10\rangle-\delta|11\rangle)_{ab}$$$$
+(a|00\rangle-b|01\rangle-c|10\rangle+d|11\rangle)_{56}\otimes$$
\begin{equation} (\alpha|00\rangle-\beta|01\rangle
-\gamma|10\rangle+\delta|11\rangle)_{ab}].
\end{equation}

 Now Bob makes an optimal POVM \cite {Bandyopadhyay} on the ancillary [17, pp.282] particles $ a and b$ to conclusively distinguish the above states.
We choose the optimal POVM in this subspace as follows:
\begin{eqnarray}
P_1=\frac {1}{x}|\Psi_1\rangle\langle\Psi_1|,P_2=\frac
{1}{x}|\Psi_2\rangle\langle\Psi_2|,     \nonumber\\
P_3=\frac{1}{x}|\Psi_3\rangle\langle\Psi_3|,P_4=\frac{1}{x}|\Psi_4\rangle\langle\Psi_4|,   \nonumber\\
P_5=I-\frac {1}{x}\Sigma_{i=1}^4|\Psi_i\rangle\langle\Psi_i|,
\end{eqnarray}
where
\begin{eqnarray}
|\Psi_1\rangle=\frac {1}{\sqrt {\frac {1}{\alpha^2}+\frac
{1}{\beta^2}+\frac {1}{\gamma^2}+\frac {1}{\delta ^2}}}
\nonumber\\
 (\frac {1}{\alpha}|00\rangle+\frac
{1}{\beta}|01\rangle+\frac {1}{\gamma}|10\rangle+\frac
{1}{\delta}|11\rangle)_{ab},
\end{eqnarray}
\begin{eqnarray}
|\Psi_2\rangle=\frac {1}{\sqrt {\frac {1}{\alpha^2}+\frac
{1}{\beta^2}+\frac {1}{\gamma^2}+\frac {1}{\delta ^2}}}
\nonumber\\
 (\frac {1}{\alpha}|00\rangle+\frac
{1}{\beta}|01\rangle-\frac {1}{\gamma}|10\rangle-\frac
{1}{\delta}|11\rangle)_{ab},
\end{eqnarray}
\begin{eqnarray}
|\Psi_3\rangle=\frac {1}{\sqrt {\frac {1}{\alpha^2}+\frac
{1}{\beta^2}+\frac {1}{\gamma^2}+\frac {1}{\delta ^2}}}
\nonumber\\
 (\frac {1}{\alpha}|00\rangle-\frac
{1}{\beta}|01\rangle+\frac {1}{\gamma}|10\rangle-\frac
{1}{\delta}|11\rangle)_{ab},
\end{eqnarray}

\begin{eqnarray}
 |\Psi_4\rangle=\frac {1}{\sqrt {\frac {1}{\alpha^2}+\frac
{1}{\beta^2}+\frac {1}{\gamma^2}+\frac {1}{\delta ^2}}}\nonumber\\
(\frac {1}{\alpha}|00\rangle-\frac {1}{\beta}|01\rangle-\frac
{1}{\gamma}|10\rangle+\frac {1}{\delta}|11\rangle)_{ab};
\end{eqnarray}
  $I$ is an identity operator;  $x$ is  a coefficient  relating  to $\alpha, \beta, \gamma, \delta$,
 $1\leq x\leq 4$, and  makes  $P_5$ to be  a positive operator.
 Obviously, we should carefully choose $x$ such  that  all the diagonal elements of $P_5$ are  nonnegative. If
the result of Bob's POVM  is $P_1$, then Bob can safely conclude
that the state of the particles 5,6 is
\begin{equation}\label{telestate}
|\Phi\rangle_{56}=(a|00\rangle+b|01\rangle+c|10\rangle+d|11\rangle)_{56}.
\end{equation}
for other result we can recover the state of particles $5,6$.By
the similar method we can make the teleportation successful in the
other outcomes of Alice's  Bell state measurement. For the sake of
saving the space we will not write them out. Evidently, when the
Bell states $|\Phi^+\rangle_{23}$ and $|\Phi^+\rangle_{14}$ are
acquired in Alice's two
 Bell state measurements,   the probability of successful teleportation
 is $\frac{F^{2}}{x}$.
 Synthesizing all  Alice's Bell state measurement cases (sixteen kinds in all), the probability of successful
teleportation in this scheme is $\frac{16*F^{2}}{x}$. Hence the
smallest $x$ corresponds to the highest probability of successful
teleportation.

\section{Conclusion}

\label{sec:alphaMZ}

\vspace{1mm}\noindent It was shown that the possibility of
teleportation of  protons as fermions ( in low energy scales) can
be focused using some  available experimental techniques. However
lastly, we have used a two-fermion particle state in one hand,
with a four-particle pure entangled state in other hand to
teleport a fermionic characteristic (spin), but it should be noted
that we have no idea on how to prepare a successful setup to
examine our suggestion. It seems that the mathematics behind it
works well. But it is honestly a long way between a mythic and
gedanken Idea and a viewable-experimental idea to have reasonable
data in laboratory.

\end{document}